\documentclass[10pt,conference]{IEEEtran}
\IEEEoverridecommandlockouts
\newcommand{\CLASSINPUTtoptextmargin}{2.4cm}
\usepackage{tabularx}
\usepackage{cite}
\usepackage{amsmath,amssymb,amsfonts}
\usepackage{algorithmic}
\usepackage{graphicx}
\usepackage{textcomp}
\usepackage{xcolor,soul,framed}
\usepackage{comment}
\usepackage{multirow}
\usepackage{hyperref}
\usepackage{fancyhdr,lipsum}

\usepackage[linesnumbered,lined,boxed,commentsnumbered,ruled,longend, noend]{algorithm2e}

\SetCommentSty{mycommfont}
\SetKwInput{KwInput}{Input}                
\SetKwInput{KwOutput}{Output}              
\def\BibTeX{{\rm B\kern-.05em{\sc i\kern-.025em b}\kern-.08em
    T\kern-.1667em\lower.7ex\hbox{E}\kern-.125emX}}

\DeclareFontFamily{U}{mathx}{}
\DeclareFontShape{U}{mathx}{m}{n}{<-> mathx10}{}
\DeclareSymbolFont{mathx}{U}{mathx}{m}{n}
\DeclareMathAccent{\widehat}{0}{mathx}{"70}
\DeclareMathAccent{\widecheck}{0}{mathx}{"71}

\fancypagestyle{mahmood}{%
   \fancyhf{} 
   
   \fancyhead[C]{979-8-3503-1090-0/23/\$31.00~\copyright~2023 IEEE. Reprinting or republishing this material for the purpose of advertising or promotion, reselling or redistributing to servers or lists, or using any copyrighted component in other works must adhere to IEEE policy. The paper has been accepted for publication at \textbf{GlobeCom 2023}, and DOI will be provided as soon as possible.}
}%

\makeatletter
\let\ps@IEEEtitlepagestyle\ps@mahmood
\makeatother

\begin{document}

\renewcommand\tabularxcolumn[1]{m{#1}}
\newcolumntype{Y}{>{\centering\arraybackslash}X}

\title{QoS-Aware Service Prediction and Orchestration in Cloud-Network Integrated Beyond 5G\\
}

\author{
    \IEEEauthorblockN{
        Mohammad Farhoudi, Masoud Shokrnezhad, and Tarik Taleb
    }
    \IEEEauthorblockA{
        Oulu University, Oulu, Finland\\
        \{mohammad.farhoudi, masoud.shokrnezhad, tarik.taleb\}@oulu.fi
    }
}

\newcommand*{\myprime}{^{\mkern 1.2mu \prime}}
\newcommand*{\mydprime}{^{\prime\prime}\mkern-1.2mu}

\maketitle

\begin{abstract}
Novel applications such as the Metaverse have highlighted the potential of beyond 5G networks, which necessitate ultra-low latency communications and massive broadband connections. Moreover, the burgeoning demand for such services with ever-fluctuating users has engendered a need for heightened service continuity consideration in B5G. To enable these services, the edge-cloud paradigm is a potential solution to harness cloud capacity and effectively manage users in real time as they move across the network. However, edge-cloud networks confront a multitude of limitations, including networking and computing resources that must be collectively managed to unlock their full potential. This paper addresses the joint problem of service placement and resource allocation in a network-cloud integrated environment while considering capacity constraints, dynamic users, and end-to-end delays. We present a non-linear programming model that formulates the optimization problem with the aiming objective of minimizing overall cost while enhancing latency. Next, to address the problem, we introduce a DDQL-based technique using RNNs to predict user behavior, empowered by a water-filling-based algorithm for service placement. The proposed framework adeptly accommodates the dynamic nature of users, the placement of services that mandate ultra-low latency in B5G, and service continuity when users migrate from one location to another. Simulation results show that our solution provides timely responses that optimize the network's potential, offering a scalable and efficient placement.
\end{abstract}




\section{Introduction}
In this fast-paced world, networking environments have evolved, leading to an increase in data flow \cite{kianpishehSurvey}. The shift in paradigm has given birth to a range of entirely new services that require rigorous Quality of Service (QoS) requirements 
\cite{ACHIR-Iot-Discovery2022}. Some of these services include the Metaverse, Unmanned Aerial Vehicles (UAVs), and Augmented Reality/Virtual Reality (AR/VR) \cite{haoyu2023xr}. With the rise of these services, ensuring reliable and efficient data flow is now essential. The QoS requirements for these services are stringent, and any delay or interruption in data flow can have severe consequences. As the evolution of networks continues to accelerate, we anticipate more novel services demanding strict QoS. 
Therefore, the need for robust and reliable networks that can handle increased data flow is more critical than ever. To meet these demands, technological advancements in network infrastructure are continuously being made. The evolution of networks has paved the way for the development of innovative solutions that cater to the QoS requirements of these novel services \cite{enableSlicingForMobility}.


Distributed edge-cloud architecture is one of the potential substrates to answer this need, which has become an indispensable part of today's computing landscape. This continuum is based on Service Oriented Architecture (SOA) and is gaining popularity due to its scalability, reliability, and availability of computing functionalities/facilities that can be used as resources. 
The purpose of edge computing is to bring data intelligence, processing, and storage closer to the network's edge, while cloud computing provides more capacity and a more reliable environment \cite{habibi2020}. Through the edge-cloud continuum environment, computer-related service requests will be answered more promptly, the quality of services will be improved, and the location of users will be tracked more accurately. 
In Beyond 5G (B5G) networks \cite{shu2022resource},
edge-cloud infrastructure is integrated into distinct domains, and Network Function Virtualization (NFV) virtualizes these resources, creating isolated virtual entities on top of physical infrastructure \cite{vnf2013}. Hence, Virtual Network Functions (VNF) and service instances are available through Software-Defined Networks (SDNs) and NFV, offering users a range of services and computing resources.

Effective service orchestration is crucial to ensure optimal service delivery, which meets both network constraints and user requirements \cite{energyEdge}. Considering it, the QoS and Quality of Experience (QoE) can be improved, and the continuity of services can be provided in an efficient manner \cite{Alshafaey2021}. One of the greatest challenges of effective service orchestration in edge-cloud computing is resource management, where the best suitable service instances should be selected for user requests, and computing and networking resources should be allocated and scheduled jointly, promoting resource sharing and maintaining a deterministic system to ensure that services and user requests are satisfied in terms of their QoS and QoE requirements, resulting in various system-level predefined objective functions, such as provider-level cost minimization (for example, through energy savings) or profit maximization (by, for instance, increasing resource utilization) \cite{li_cognitive_2021}.

As of now, different concepts, architectures, and paradigms have been considered in the approaches proposed for service orchestration. Zhang \textit{et al.} \cite{zhang2019} have developed an adaptive interference-aware heuristic approach to optimize VNF placement, which has been shown to effectively handle traffic variation and improve the total throughput of accepted requests. Li \textit{et al.} \cite{LI2020} have presented a resource management and replica allocation strategy for edge-cloud computing systems, which aims to reduce financial costs while maintaining performance and data consistency. Additionally, 
a heuristic near-optimal solution to the joint problem of networking and computing resource allocation for 5G networks was presented \cite{shokrnezhad2023}. This work proposed an optimal approach to find the optimal solution to the joint problem.
Dant \textit{et al.} \cite{Dang2021} have presented the architecture of SDNized Information-Centric Networking (ICN) technologies which incorporate service placement.

Although the proposed methods in these studies are effective in addressing the resource allocation problem, their applicability to real-world scenarios remains a challenge, as they fail to cater to the dynamic nature of users and their requests, making service continuity a challenge. These approaches provided static allocation which will be useless in applications like the Metaverse whereby users and requests are subject to changes on a millisecond basis. Moreover, in most of the previous works, resource allocation has been isolated to the cloud domain, and the network is solely viewed as a pipeline with no cognitive ability to adapt regarding the changes in the system. Clearly, such solitary approaches are ineffective due to disregarding interdependencies among various domains and resources. A failure in one domain can have far-reaching effects on the other ones, so orchestrating services from a siloed perspective may not be adequate to achieve the desired system performance.


This study aims to address the existing gap in the literature by examining the joint problem of service instance placement and assignment, as well as path selection in the context of an edge-cloud continuum environment wherein users are moving and requests are changing their Point of Attachment (PoA) over time. To address this problem, the first step is to predict which requests will arrive at each PoA in the near future, followed by a joint assignment of networking and computing resources to meet their requirements. In particular, we consider capacity limitations of the resources, and End-to-End (E2E) delays, with the goal of minimizing total cost.
Our main contributions to this paper are:

\begin{itemize}
    \item Formulating the joint problem of service placement and resource allocation in the edge-network-cloud integrated infrastructure as a Mixed Integer Non-Linear Programming (MINLP) problem.
    \item Proposing a deep reinforcement learning method for predicting the arrival point of requests utilizing historical data for smoother handling of user dynamicity and improving service continuity.
    \item Devising a novel heuristic approach based on the water-filling algorithm to identify near-optimal solutions for the placement of service instances on edge-cloud nodes and allocating networking resources regarding the QoS requirements of requests, utilizing the output of the learning method to minimize delay and cost, resulting in the more efficient placement of resources.
\end{itemize}

The upcoming sections of this paper are arranged coherently as follows. Commencing with Section \ref{sec:system_model}, the system model is outlined, followed by a detailed resource allocation problem formulation in Section \ref{sec:problem}.
The heuristic approaches are then presented in Sections \ref{sec:method} with utmost clarity, so as to easily comprehend the technical aspects of the proposed method of service prediction and orchestration. In Section \ref{sec:results}, the numerical results are illustrated with the help of appropriate figures, thereby providing a clear understanding of the research outcomes. Finally, Section \ref{sec:conclusion} offers concluding remarks and future directions that encapsulate the study's findings.

\section{System Model} \label{sec:system_model}
In the following, the system model is provided. This paper examines three main components of the system: edge-cloud infrastructure, service providers, and user requests. 

\subsection{Edge-cloud Infrastructure}
The edge-cloud infrastructure consists of a network that connects computing resources available for deploying instances of services. The network, denoted by $\mathcal{G(\boldsymbol{\mathcal{N}}, \boldsymbol{\mathcal{L}}, \boldsymbol{\mathcal{P}})}$, consists of two domains (i.e., access and core), where $\boldsymbol{\mathcal{N}}$ is the set of edge-cloud nodes with size \( \mathcal{N} \), $\boldsymbol{\mathcal{L}} \subset \{ l : (n, n\myprime)|n, n\myprime \in \mathcal{N} \}$ is the set of links with size \( \mathcal{L} \), and $\boldsymbol{\mathcal{P}} = \{ p: (\mathcal{H}_p, \mathcal{T}_p) | p \subset \boldsymbol{\mathcal{L}} \}$ represents the set of directional paths with size \( \mathcal{P} \). Each path $p$ is determined by its head node (\( \mathcal{H}_p \)) and tail node (\( \mathcal{T}_p \)), and $\mathcal{J}_{p,l}$ is a binary parameter equal to $1$ if path $p$ contains link $l$. As each edge-cloud node is equipped with computing resources, it can be considered a host for deploying service instances. It is predetermined that the computing resources available on each node are limited by a predefined capacity threshold \( \widehat{\mathcal{C}}_n \), and the bandwidth available on each link is limited by a corresponding capacity \( \widehat{\mathcal{L}}_l \). The cost of using each node and a link is associated with a corresponding cost, denoted by \( \mathcal{\overline{C}}_n \) and \( \mathcal{\overline{L}}_l \). Note that the network is structured at different levels, and the nodes are distributed so that the more nodes close to the cloud, the higher the capacity and lower the costs. Thus, nodes near end-users or entry points have expensive but limited computing resources, while central nodes have cheaper and higher capacity computing resources \cite{kianpishehSurvey}.
    
\subsection{Service Providers}
Participating in the system are $\mathcal{S}$ service providers, each of which offers a set of service instances $\boldsymbol{\mathcal{I}}_s = \{1,2,...,\mathcal{I}_s\}$ with size $\mathcal{I}_s$. Consequently, the set of services is represented by $\boldsymbol{\mathcal{S}}=\{ \boldsymbol{\mathcal{I}}_1, \boldsymbol{\mathcal{I}}_2, ..., \boldsymbol{\mathcal{I}}_S \}$. Although each instance is capable of handling multiple requests, its capacity is limited by a predetermined threshold \( \widehat{\mathcal{I}}_{s,i} \), and the cost of using each service instance is denoted by \( \mathcal{\overline{I}}_{s,i}\). 

\subsection{User Requests}
The system contains a set of \( \mathcal{R} \) active requests, denoted by $\boldsymbol{\mathcal{R}}$, where each request $r$ arrives in the system at time $\mathcal{T}_r$ and continuously demands a service $\mathcal{S}_r$ to send its inquiry traffic to one of its instances for a particular operation and then receives the response. Users exhibit dynamic behavior in the system by changing their locations over time. For this reason, \( \mathcal{E}^{t}_{r} \) identifies the node (PoA) from which each request originates at each time slot. Upon reaching the PoA, the most appropriate service instance should be selected for request $r$ based on its requirements such as the minimum service capacity $\widecheck{\mathcal{I}}^{t}_{r}$, minimum network bandwidth $\widecheck{\mathcal{L}}^{t}_{r}$, maximum acceptable E2E delay $\widecheck{\mathcal{D}}^{t}_{r}$, traffic burstiness $\widecheck{\mathcal{B}}^{t}_{r}$, maximum packet size $\widecheck{\mathcal{Z}}^{t}_{r}$, and $\widecheck{\mathcal{O}}_{r}$ indicating the upper limit of overall E2E delay that can be tolerated by request $r$ over $\mathcal{T}$ time slots, also known as the Service-Level Agreement (SLA) requirement. 

\section{Problem Definition} \label{sec:problem}
In this section, the joint problem of resource allocation is defined as an MINLP formulation, taking into account instance placement and assignment (\ref{subsec: placement}), path selection (\ref{subsec: path}), and delay constraints (\ref{subsec: sla}) with the aim of minimizing the overall cost (\ref{subsec: objective}) to ensure that QoS requirements of requests are continuously met at the lowest possible cost, given that they are changing their PoA over time.


\subsection{Objective Function} \label{subsec: objective}
This objective function (OB) seeks to minimize the total cost of allocated resources over the time interval $\boldsymbol{\mathcal{T}}$ (beginning at time $1$ and ending at time $\mathcal{T}$).  Specifically, this equation captures the cost associated with the assignment of requests to instances, the placement of instances on edge-cloud nodes, and the selection of inquiry and response paths for requests. $\Ddot{\mathcal{A}}^{t}_{r, i}$ and $\Ddot{\mathcal{E}}^{t}_{i,n}$ are binary variables that indicate the instance of request $r$ (considering that the instance of request $r$ must be chosen from among the instances of $\mathcal{S}_r$) and the host node of instance $i$ respectively at time $t$, and $\Ddot{\mathcal{L}}^{t}_{r}$ is a continous variable that shows the total cost of allocated paths to request $r$ at time $t$. In this equation and what follows, $i$ is iterating over the instances of $\mathcal{S}$. 

\footnotesize
\begin{equation*}\label{OBJECTIVEFUNCTION1}\sum_{\boldsymbol{\mathcal{T}}, \boldsymbol{\mathcal{N}}, \boldsymbol{\mathcal{S}}} \Ddot{\mathcal{E}}^{t}_{i,n} \overline{\mathcal{C}}_n + \sum_{\boldsymbol{\mathcal{T}}, \boldsymbol{\mathcal{S}}, \boldsymbol{\mathcal{R}}}\Ddot{\mathcal{A}}^{t}_{r,i} \overline{\mathcal{I}}_{s,i} + \sum_{\boldsymbol{\mathcal{T}}, \boldsymbol{\mathcal{R}}} \Ddot{\mathcal{L}}^{t}_{r} \tag{OB}
\end{equation*}
\normalsize

\subsection{Instance Placement and Assignment Constraints} \label{subsec: placement}
The first step is to ensure that each request is always assigned to a single instance of the service (C1). C2 ensures that each service instance selected by at least one request is placed on at least one available edge-cloud node at the time requested. To avoid congestion and ensure the framework's reliability, the total number of requests assigned to each service instance cannot exceed the capacity of the instance in each time slot (C3). Nodes are only able to handle a limited capacity as well (C4).

\vspace{-10pt}
\footnotesize
\begin{align*}\label{placement_constraints}
    & \footnotesize  \sum_{\boldsymbol{\mathcal{S}}} \Ddot{\mathcal{A}}^{t}_{r,i} = 1 \quad \forall r \in \boldsymbol{\mathcal{R}}, t \in [\mathcal{T}_r,\mathcal{T}] \tag{C1}
    \\
    & \sum_{\boldsymbol{\mathcal{N}}} \Ddot{\mathcal{E}}^{t}_{i,n} > \left( \sum_{\boldsymbol{\mathcal{R}}} \Ddot{\mathcal{A}}^{t}_{r,i} \right) / \mathcal{R} \quad \forall i \in \boldsymbol{\mathcal{S}}, t \in [\mathcal{T}_r,\mathcal{T}]   \tag{C2}
    \\
    &\sum_{\boldsymbol{\mathcal{R}}} \Ddot{\mathcal{A}}^{t}_{r,i} \widecheck{\mathcal{I}}^{t}_{r} \leq \widehat{\mathcal{I}}_{s,i} \quad  \forall i, t \in \boldsymbol{\mathcal{S}}, \boldsymbol{\mathcal{T}} \tag{C3}
    \\
    &\sum_{\boldsymbol{\mathcal{S}}, \boldsymbol{\mathcal{R}}}\Ddot{\mathcal{E}}^{t}_{i,n} \Ddot{\mathcal{A}}^{t}_{r,i} \widecheck{\mathcal{I}}^{t}_{r} \leq \widehat{\mathcal{C}}_n \quad \forall n, t \in \boldsymbol{\mathcal{N}}, \boldsymbol{\mathcal{T}} \tag{C4} 
\end{align*}
\normalsize

\subsection{Path Selection Constraints} \label{subsec: path}
In order to deliver inquiry traffic of a request to its assigned instance and return the response, it is necessary to assign a feasible E2E route for each request within the specified time slot (C5 and C6). To do so, a unique inquiry path is selected for each request, originating from its entry node (PoA) and concluding at the chosen service instance, denoted by $\overrightarrow{\mathcal{R}}^{t}_{r,p}$. For each request, the corresponding response path, or $\overleftarrow{\mathcal{R}}^{t}_{r,p}$, is also determined using a similar approach, but with the order of the nodes reversed. In other words, the response path starts at the selected service instance and ends at its PoA. Additionally, a capacity limitation applies to the number of requests assigned to each path at any given time (C7), and C8 computes the total path allocation cost for each request.

\vspace{-10pt}
\footnotesize
\begin{align*}\label{path_selection_constraints}
    & \footnotesize \sum_{\boldsymbol{\mathcal{P}} | \mathcal{H}_p=\mathcal{E}^{t}_{r} \& \mathcal{T}_p = n } \overrightarrow{\mathcal{R}}^{t}_{r,p} = \sum_{\boldsymbol{\mathcal{S}}}\Ddot{\mathcal{A}}^{t}_{r,i} \Ddot{\mathcal{E}}^{t}_{i,n} \quad \forall r, n, t \in \boldsymbol{\mathcal{R}}, \boldsymbol{\mathcal{N}}, \boldsymbol{\mathcal{T}} \tag{C5}
    \\
    & \sum_{\boldsymbol{\mathcal{P}} | \mathcal{H}_p=n \& \mathcal{T}_p = \mathcal{E}^{t}_{r} } \overleftarrow{\mathcal{R}}^{t}_{r,p} = \sum_{\boldsymbol{\mathcal{S}}}\Ddot{\mathcal{A}}^{t}_{r,i} \Ddot{\mathcal{E}}^{t}_{i,n} \quad \forall r, n, t \in \boldsymbol{\mathcal{R}}, \boldsymbol{\mathcal{N}}, \boldsymbol{\mathcal{T}} \tag{C6}
    \\
    &\sum_{\boldsymbol{\mathcal{R}}} \widecheck{\mathcal{L}}^{t}_{r} \left( \sum_{\boldsymbol{\mathcal{P}}} \mathcal{J}_{p,l} \left( \overrightarrow{\mathcal{R}}^{t}_{r,p} + \overleftarrow{\mathcal{R}}^{t}_{r,p} \right) \right) \leq \widehat{\mathcal{L}}_{l} \quad \forall l, t \in \boldsymbol{\mathcal{L}}, \boldsymbol{\mathcal{T}} \tag{C7}
    \\
    &\Ddot{\mathcal{L}}^{t}_{r} = \sum_{\boldsymbol{\mathcal{L}}} \overline{\mathcal{L}}_l \left( \sum_{\boldsymbol{\mathcal{P}}} \mathcal{J}_{p,l} \left( \overrightarrow{\mathcal{R}}^{t}_{r,p} + \overleftarrow{\mathcal{R}}^{t}_{r,p} \right) \right) \quad \forall r, t \in \boldsymbol{\mathcal{R}}, \boldsymbol{\mathcal{T}} \tag{C8}    
\end{align*}
\normalsize

\subsection{Service-Level Agreement} \label{subsec: sla}
The final set of constraints focuses on delay limitations, with the objective of fulfilling the QoS requirements of users. $\mathcal{D}^{t}_{r,l}$ and $\mathcal{D}^{t}_{r}$ are continuous variables that represent the delay of link $l$ and E2E delay respectively (including network delay and computing delay) for request $r$ at time $t$ \cite{shokrnezhad2022} while considering the number of priority is equal to 1. During each time period, it is necessary to ensure that the maximum acceptable delay for requests is maintained (C11). Furthermore, the sum of delays experienced by the requests of each user across all time slots should not exceed its SLA requirement (C12).

\vspace{-10pt}
\footnotesize
\begin{align*}\label{SLA_constraints}
    & \mathcal{D}^{t}_{r,l} = \left( \sum_{\boldsymbol{\mathcal{R}} | r\myprime \neq r} \widecheck{\mathcal{B}}^{t}_{r'} +\widecheck{\mathcal{Z}}^{t}_{r'} \right) / \widehat{\mathcal{L}}_{l} \quad \forall r, l, t \in \boldsymbol{\mathcal{R}}, \boldsymbol{\mathcal{L}}, \boldsymbol{\mathcal{T}} \tag{C9} 
    \\
    & \mathcal{D}^{t}_{r} = \sum_{\boldsymbol{\mathcal{P}}, \boldsymbol{\mathcal{L}}} \mathcal{J}_{p,l} \mathcal{D}^{t}_{r,l} \left( \overrightarrow{\mathcal{R}}^{t}_{r,p} + \overleftarrow{\mathcal{R}}^{t}_{r,p} \right) + \widecheck{\mathcal{Z}}^{t}_{r}/\widecheck{\mathcal{I}}^{t}_{r} \quad \forall r, t \in \boldsymbol{\mathcal{R}}, \boldsymbol{\mathcal{T}} \tag{C10}
    \\
    & \mathcal{D}^{t}_{r} \leq \widecheck{\mathcal{D}}^{t}_{r} \quad \forall r, t \in \boldsymbol{\mathcal{R}}, \boldsymbol{\mathcal{T}} \tag{C11}
    \\
    & \sum_{\boldsymbol{\mathcal{T}} | t \geq \mathcal{T}_{r}} \mathcal{D}^{t}_{r} \leq \widecheck{\mathcal{O}}_{r} \quad \forall r \in \boldsymbol{\mathcal{R}} \tag{C12}
\end{align*}
\normalsize

\subsection{Problem}\label{section:problem}
After thorough consideration of the relevant constraints and identification of the overall objective function, the problem of qos-Aware Service orChEsTration In edge-Cloud (ASCETIC) is formulated as follows:

\footnotesize
\begin{align}\label{EQ_ASCETIC}
    \text{ASCETIC: } \textit{ min } \text{OB} \textit{ s.t. } \text{C1 - C12.} 
\end{align}
\normalsize

\section{Proposed method} \label{sec:method}
The Problem \ref{section:problem} is classified as NP-hard.
This has been demonstrated through the reduction of the multidimensional knapsack problem 
elaborated in \cite{faticanti_cutting_2018}. Consequently, in the worst-case scenario, the complexity of solving this problem could increase to the extent of the solution space size \cite{pataki_basis_2010}. To determine the optimal allocation for a given request, each node, instance, and path must be evaluated at least once. Since allocating resources for any request at any time slot affects and is affected by allocations for others, all possible sequences of requests and time slots must be considered, yielding a solution space of size $\mathcal{R}!\mathcal{T}!\mathcal{N}\mathcal{S}\mathcal{P}^2$. As a result, identifying the optimal solution for large-scale instances in a timely manner is impractical, even with all the necessary information available and a fully-aware environment. Further complicating the situation is the fact that in a continuous network, not all the required information (such as the requests for future time slots and their PoA) are accessible. To conquer knowledge imperfectness/inadequacy and lead a quality result in a timely manner, an approach named Water-fIlling of Service placEment (WISE) is proposed in Algorithm \ref{WISE} involving two separate sections: prediction (steps 2-15) and orchestration (steps 16 - 38). These mechanisms iterate for each time slot.

\begin{algorithm}[t!]
\caption{WISE}
\label{WISE}
\KwInput{$\mathcal{T}$, $\epsilon$, $\epsilon'$, $\widetilde{\epsilon}$, $\theta_0 \gets \{\}$, and $\alpha_0 \gets \{\}$}
\For{each $\tau$ in $[1:\mathcal{T}]$}
{
    update $\theta_{\tau}$ using the arrived requests at the PoA\\
    \If{$\tau < m$}
    {
        $\alpha_{\tau+1} \gets $ select a set of $z$ random services
    }
    \Else
    {
        $\zeta \gets$ generate a random number from $[0:1]$ \\
        \If{$\zeta > \epsilon$}
        {
            $\alpha_{\tau+1} \gets$ select $z$ services with top Q values
        }
        \Else
        {
            $\alpha_{\tau+1} \gets$ select a set of $z$ random services
        }
        calculate $\rho_{\tau}$ \\
        $mem \gets mem \cup \{(\theta_{\tau-1}, a_{\tau-1}, \rho_{\tau},\theta_{\tau})\}$ \\
        choose a sample form $mem$ and train the agent \\
        \If{$\epsilon > \widetilde{\epsilon}$}
        {
            $\epsilon \gets \epsilon - \epsilon'$
        }
    }
    $\boldsymbol{\alpha} \gets$ collect $\alpha_{\tau+1}$ of all PoAs \\
    convert $\boldsymbol{\alpha}$ to a (Requests, PoAs) table \\
    \While{$\boldsymbol{\mathcal{R}}$ is not empty}
    {
        $r \gets$ the tightest E2E delay requirement request \\
        $\boldsymbol{\eta} \gets$ the set of PoAs requesting $\mathcal{S}_r$\\
        $\mathcal{D} \gets \infty$ \\
        \For{each $n_1 \in \boldsymbol{\mathcal{N}}$}
        {
            $\mathcal{D}_{n_{1}} \gets 0$ \\
            \For{each $n_2 \in \boldsymbol{\eta}$}
            {
                $\boldsymbol{\mathcal{P}}_1 \gets$ the set of paths from $n_1$ to $n_2$ \\
                $p_1 \gets$ $p \in \boldsymbol{\mathcal{P}}_1$ with the lowest delay\\
                $\boldsymbol{\mathcal{P}}_2 \gets$ the set of paths from $n_2$ to $n_1$ \\ 
                $p_2 \gets$ $p \in \boldsymbol{\mathcal{P}}_2$ with the lowest delay\\
                $\mathcal{D}_{p} \gets$ calculate delay + cost for $p_1 + p_2$ \\
                $\mathcal{D}_{n_{1}} \gets \mathcal{D}_{n_{1}} + \mathcal{D}_{p} + \mathcal{\overline{C}}_{n_1}$
            }
            \If{$\mathcal{D}_{n_{1}} < \mathcal{D}$}
            {
                $n \gets n_{1}, \overrightarrow{p} \gets p_1, \overleftarrow{p} \gets p_2$
            }
        }
        \While{$n$ is feasible}
        {
            Place a new instance $i$ of $\mathcal{S}_r$ on $n$ \\
            \While{$i$ is feasible}
            {
                \For{$r' \in \boldsymbol{\mathcal{R}}$}
                {
                    $\Ddot{\mathcal{A}}^{t}_{r',i} \gets 1, \overrightarrow{\mathcal{R}}^{t}_{r',\overrightarrow{p}} \gets 1, \overleftarrow{\mathcal{R}}^{t}_{r',\overleftarrow{p}} \gets 1$ \\
                    remove $r'$ from $\boldsymbol{\mathcal{R}}$
                }
            }
        }
    }
}
\end{algorithm}

The prediction mechanism focuses on determining the next PoA of each request to adjust the allocated resources apriori to maintain the continuity of service provision. To do so, each PoA (through steps 2 to 15) employs a Double Deep Q-Learning (DDQL) agent at each point in time wherein Recurrent Neural Networks (RNNs) are used to approximate the likelihood that each request $r$ will be requested at the next time (Q values). The agent's state ($\theta$) is the vector of arrived requests to this PoA during the last $m$ time slots, the action ($\alpha$) returns the list of $z$ requests with the highest likelihood, and the reward ($\rho$) is the number of requests predicted correctly. Note that the action in each iteration is chosen by the $\epsilon$-greedy policy that follows the evaluation function of the corresponding agent with probability $(1-\epsilon)$ and chooses a random action with probability $\epsilon$. During the training process, the probability decreases linearly from $\epsilon$ to $\widetilde{\epsilon}$. Besides, to improve the efficiency, the observed transitions are stored in a memory bank ($mem$), and the neural network is updated by randomly sampling from this pool \cite{mnih_human-level_2015}.

The orchestration mechanism is dedicated to determining the most appropriate allocations for the predicted requests on available nodes, paths, and instances. After collecting the expected requests from all PoAs in a central controller, it is first necessary to transform them into a PoA requests table. The algorithm then proceeds to iterate through each request $r$, beginning with the request with the most demanding time requirement. Then, a node is selected with the minimum overall delay and cost to all PoAs predicting to have requests demanding the same service as request $r$. If the selected node is feasible in terms of E2E delay and computing capacity requirements, new instances will be located on it, and then requests with the same target service will be assigned to these instances. This operation will be continued till no more instances can be added to this node, so a new node will be selected based on the arrival point of the remaining requests, and the algorithm will be continued till all requests are investigated. Note that WISE has a worst-case complexity of $O(\mathcal{T}\mathcal{R}\mathcal{N}^2\mathcal{P}^2)$, since at each time, it investigates $\mathcal{R}$ requests, and on each iteration, it checks inquiry and response paths between all nodes and the list of PoAs.

\section{Simulation Results}\label{sec:results}

The purpose of this section is to examine the efficiency of the WISE method numerically by considering the cost of consuming nodes, instances, and links, as well as the E2E delay, and the number of supported requests (as a metric for assessing service continuity). WISE is compared to various approaches such as finding the optimal solution through solving (\ref{EQ_ASCETIC}) using CPLEX, selecting instances and nodes randomly to meet requests, and implementing a service placement and discovery method as described in \cite{Dang2021} on Connected and Cooperative Autonomous Mobility (CCAM). The simulation parameters are enumerated in Table \ref{table:simulation_paramter}. In so far, as the issue remains viable, the residual parameters can be selected in a flexible manner. Due to the inherent variations in parameters such as $\mathcal{\overline{L}}_l$ and $\mathcal{\overline{C}}_n$, it is reasonable to expect fluctuations across the costs of all methods.


As part of our evaluation procedure, we alter the number of edge-cloud nodes and requests in the system to determine the effect of these changes on the provided solutions. Considering future applications (such as Internet of Things (IoT) use cases for building smart cities \cite{Elmarai2021}, the Metaverse multiverses \cite{haoyu2023xr}, and UAV-based surveillance and delivery scenarios \cite{Motlagh2016}) where a massive amount of real-time data with stringent QoS requirements must be collected and processed, the B5G infrastructure size is expected to increase, including a large number and vast variety of networking and computing resources integrated from edge to cloud. In addition, the system may experience sudden spikes in the number of active requests when these applications are fully realized and implemented. Therefore, it is beneficial to validate the algorithm with varying numbers of nodes and requests to ensure that the system is scalable and able to provide a satisfying user experience.

\begin{table}[t!]
\caption{Simulation Parameters}
\vspace{-20pt}
\label{table:simulation_paramter}
\begin{center}
\begin{tabular}{|c|c|}
\hline
\textbf{Parameter} & \textbf{Value} \\
\hline
\multirow{2}{*}{number of links ($\mathcal{L}$)} & $\sim \mathcal{U}\{3\mathcal{N}, 5\mathcal{N}\}$, where the \\ & resulted graph is connected. \\
number of services ($\mathcal{S}$) & $20$ \\
number of instances for each service ($\mathcal{I}_s$) & $5$ \\
state size of agents ($m$) & $100$ \\
action size of agents ($z$) & $\{3, 10\}$ \\
cost of each link ($\mathcal{\overline{L}}_l$) & $\sim \mathcal{U}\{10, 20\}$ \\
cost of each node ($\mathcal{\overline{C}}_n$) & $\sim 50^{\; \mathcal{U}(\alpha, \alpha+1)}$ \\
cost of each instance ($\mathcal{\overline{I}}_{s,i}$) & $\sim 20^{\; \mathcal{U}(\alpha, \alpha+1)}$ \\
capacity of each link ($\widehat{\mathcal{L}}_l$) & $\sim \mathcal{U}\{100, 150\}$ Gbps \\
capacity of each node ($\widehat{\mathcal{C}}_n$) & $\sim 50 \; \mathcal{U}(\alpha, \alpha+1)$ Gbps \\
capacity of each instance ($\widehat{\mathcal{I}}_{s,i}$) & $\sim 20 \; \mathcal{U}(\alpha, \alpha+1)$ Gbps \\
\hline
\end{tabular}
\end{center}
\vspace{-0.7cm}
\end{table}

\begin{figure*}[t!]\centering
\vspace{0.4cm}
\includegraphics[width=7.1in]{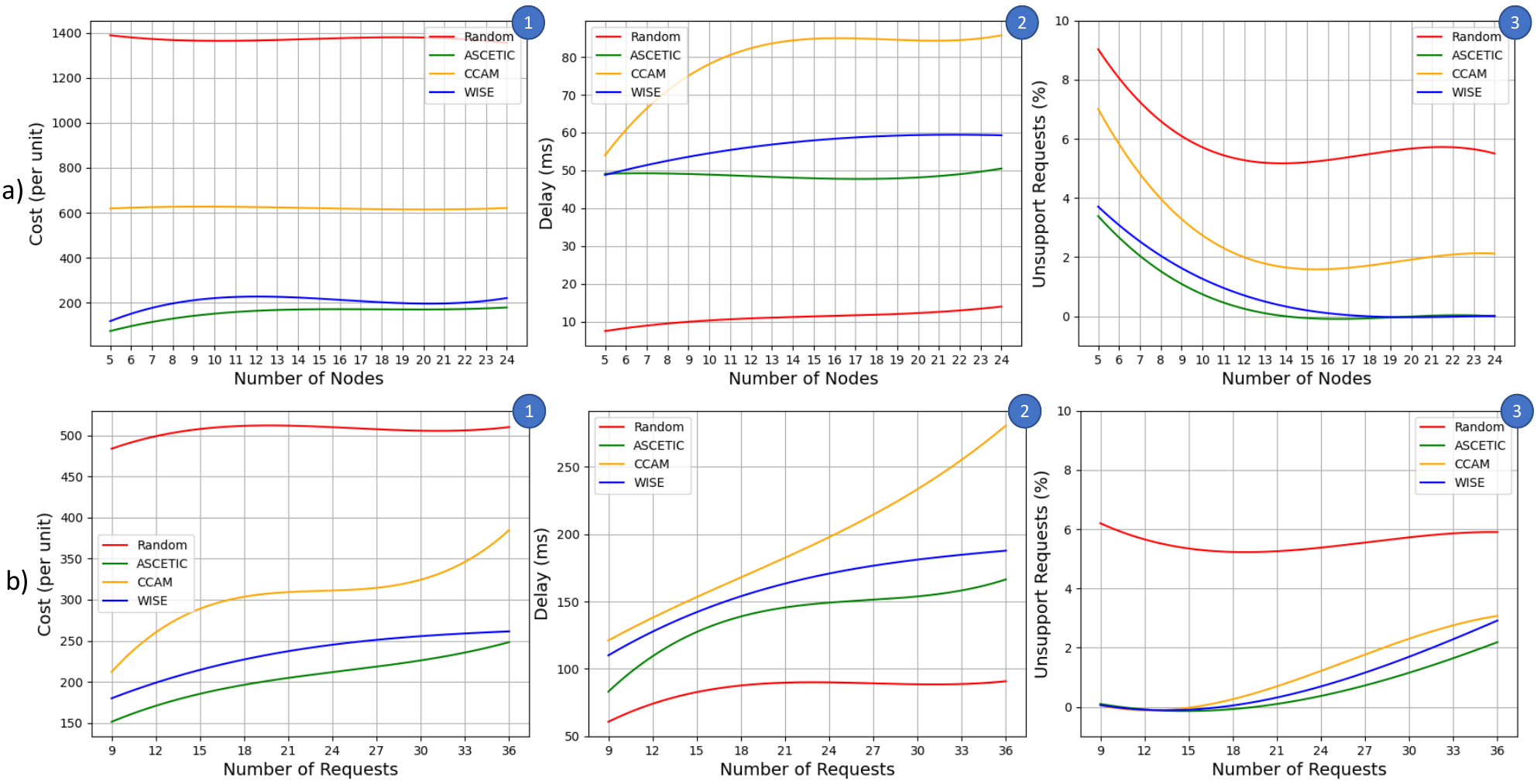}
\vspace{-0.55cm}
  \caption{(1) cost, (2) E2E delay, and (3) the number of unsupported requests for WISE vs. the ASCETIC, random selection, and CCAM methods \cite{Dang2021} when (a) the number of nodes increases, and (b) the number of requests increases}
    \vspace{-0.55cm}
    \label{figure:simulation_results}
\end{figure*}

Figure \ref{figure:simulation_results} presents the results, depicting the variations in the cost and E2E delay of allocated resources, as well as the total number of unsupported requests with increasing numbers of nodes in (a) and requests in (b). 
Notably, even in the optimal solution, the cost and delay are subject to change due to multiple factors, including changes in PoAs over time; variation in nodes, instances, and link capacities; and shifts in the minimum required capacity and bandwidth for requests.

Furthermore, the values indicated on unsupported requests represent the average of multiple runs on the system.

The sub-figures in Figure \ref{figure:simulation_results} illustrate the superior performance of the WISE approach compared to the CCAM method. Serving instances and requests for a single service with a single node is the primary drawback of the CCAM method. When the network contains a small number of nodes, i.e., when computing resources are closer to PoAs, the CCAM method performs adequately in terms of minimizing delay. However, the E2E delay increases when the number of nodes increases and high-capacity nodes are located far from entry points. In addition, it lacks in several areas, including the cost of path selection to reach particular nodes. Isolating the provisioning of each service to a single node in CCAM also results in an inability to handle all requests as the number of user requests (PoAs) grows. 

Similarly, the random method is less efficient than WISE, regardless of the number of requests and nodes. This method involves randomly placing each instance on network nodes without taking into account the ever-changing nature of users; as a result, the number of supported requests is insufficient and service continuity is deteriorating. Besides, this approach incurs high costs each time it is employed, and despite having a low delay with a small number of nodes, it frequently fails to fulfill requests. It is imperative to note that only the delay of supported requests is considered in subfigures 2; thus, it is reasonable to observe samples where the WISE method, which supports all requests, exhibits longer delays than the random method, which does not support all requests.

In terms of service placement and resource allocation, WISE exhibits an average total cost exceeding 91\% of the optimal, regardless of the size of the network, and a delay within the desired range for users' SLA. This indicates that the WISE algorithm can place services and allocate resources in a near-optimal manner, even in large networks. It is noteworthy that the average cost and delay remain significantly low regardless of the number of requests or the number of nodes. In spite of this, the delay is slightly swollen as the number of requests increases and the problem becomes more complicated due to a large number of nodes and links. Meanwhile, WISE is capable of timely responses and can place services and instances appropriately, while it is impossible to find the optimal solution to the ASCETIC problem in a reasonable amount of time. Accordingly, WISE demonstrates that it is efficient in placing services and allocating resources for large numbers of requests, is low in latency and cost, provides timely responses, and ensures service continuity compared to other approaches.

\section{Conclusion}\label{sec:conclusion}
In this study, we addressed the challenges of providing reliable and efficient service continuity in dynamic and ever-changing systems, particularly in the context of edge-cloud infrastructures for B5G networks. An MINLP problem of service placement and resource allocation in a network-cloud continuum environment, while accounting for capacity constraints, changing user behavior, and link and E2E delays, was first formulated with the objective of minimizing overall costs. Next, we proposed a water-filling-based algorithm empowered by a DDQL-based technique leveraging RNNs to solve the NP-hard problem. Simulation results demonstrated that our proposed approach is scalable, efficient, and reliable enough to be used in real-world use cases because it accommodates continuity of services when users move from one location to another and the placement of services that require extremely low latency. As a potential future direction, we plan to consider users with dynamic QoS requirements and resources with dynamic capacities and energy consumptions over time.


\section*{Acknowledgment}
This research work is partially supported by the Business Finland 6Bridge 6Core project under Grant No. 8410/31/2022, the Research Council of Finland (former Academy of Finland) IDEA-MILL project under Grant No. 352428, the European Union’s Horizon Europe research and innovation programme under the 6GSandbox project with Grant Agreement No. 101096328, and the Research Council of Finland 6G Flagship Programme under Grant No. 346208.


\vspace{-0.25cm}
\bibliographystyle{IEEEtran}
\bibliography{main}

\end{document}